\documentstyle[prb,aps]{revtex} 
\begin{document}
\parindent0em
\draft
\twocolumn[\hsize\textwidth\columnwidth\hsize\csname @twocolumnfalse\endcsname
\title{  
Fluctuators in disordered metallic point contacts: a simulation   
approach} 
\author{V.I. Kozub\cite{VIKaddr} and C. Oligschleger\cite{COaddr}} 
\address{Institut f\"ur Festk\"orperforschung, Forschungszentrum J\"ulich,\\
D-52425 J\"ulich, Germany} 
\date{\today}
\maketitle 
\begin{abstract}
Electron transport through amorphous monatomic metallic
structures generated earlier by molecular dynamics simulations is studied
numerically. The interference of electronic trajectories
backscattered by the structural disorder probes the 
multistable structural relaxations    
responsible for low-frequency noise in real metallic contacts.
The structure of these modes is visualized;
the dependence of noise magnitude on size and structure of
the modes is studied. 
The transition from
multistable behaviour to a more complex one is observed for
temperatures far below the melting
temperature.
The current fluctuations observed numerically
resemble the complex behaviour reported earlier for current
noise in small metallic structures at moderate temperatures.
\end{abstract}
]
\section{Introduction}
In recent years substantial progress in the study of the flicker 
noise problem was achieved. This is mainly connected with 
experiments on small conducting systems, where the $1/f$ noise
was decomposed \cite{1,2} into a sum of contributions due to ``elementary''
defects with internal degrees of freedom - ``fluctuators''.
These defects switch between two possible configurations and cause  
the ``telegraph'' resistance noise in small contacts. 
In metallic
systems they are believed to be related to structural defects which 
at low temperatures are seen as the well-known two-level systems   
\cite{3,4,5,6,7,8}.
We would also like to mention works where the two-level fluctuation
studied exists in atomic size contacts produced with help of the break-junction
method \cite{Muller}.  
However, there is still no clear understanding of the microscopic 
nature of these defects. 
 
Many experimental results can be explained \cite{9} by the model of
independent fluctuators described by rate equations for the occupation
numbers. The soft-potential approach which relates the fluctuators
to some ``soft'' double-well effective potentials,
has been used to describe this picture \cite{8,9}. It fails, however, to give
a microscopic insight into the problem. On the other hand, some
experiments revealed a much more complex behaviour exhibiting in
particular fluctuator interactions and non-stationary chaotic
behaviour of the fluctuator parameters \cite{2}. 
These features can not easily be described within the framework of
the soft-potential 
model. Thus a microscopic analysis of the problem would be 
desirable. However, a description of disordered 
materials based on  purely analytical methods  
meets obvious difficulties. Numerical simulations
based on molecular dynamics (MD) methods 
are used successfully to model the structure of glassy materials
as well as their thermodynamic properties including the density of states.  
In particular, many details related to low-energy 
vibrational excitations in these materials have been revealed this 
way \cite{LS}.
To our knowledge, electron transport through disordered 
media has not been treated by these
numerical methods.

Real metallic point contacts studied
experimentally can be as small as $3 - 10$ nm in diameter and thus consist 
of only $\sim 10^3-10^4$ atoms which is just of the order accessible 
to numerical simulations. Moreover, the structure of these
constrictions is not expected to be perfectly crystalline
especially in view of the amorphous films used to prepare these
point contacts.
  
Therefore, it is attractive to use a simulation approach to study 
slowly relaxing metastable configurations in disordered metallic structures
with an application to low-frequency noise. In the following we
will give the results of a first such study of noise within the framework
of simplified model considerations. 

The main results of our paper described in the following are: 
1) We have identified local
atomic arrangements leading to metastable configurations. 
2) A transition
from a simple telegraph signal to a more complex behaviour
with increasing temperature was observed which reproduces the
experimental behaviour; it is important that the ratio
of this ``critical'' temperature to the melting temperature 
agrees with experiment.
3) The magnitude of the resistance fluctuations and its dependence
on the number of atoms forming the ``fluctuator'' is considered;
the results of the simulations are in agreement with experiment.

\section{Theoretical aspects} 
Our aim is to analyze numerically charge transport 
through disordered structures prepared previously by  
molecular dynamics \cite{SOL}. The structures studied are considered 
to model a 3D cubic microbridges (with side-length $a$) 
between bulk contacts. 
The atoms forming the 
structures are taken to be scatterers for the incident 
electrons. We consider  the weak scattering limit 
where the mean free path of the electrons, $l$, is larger 
than $a$. This implies that the scattering cross-section $s$ obeys  
the inequality $n_i s < a^{-1}$ where $n_i = N/a^3$ and $N$ is the
total number of atoms in the structure. In real
glassy metals the mean free path is significantly larger
than the electron wavelengths, and therefore, the approximation is 
justified for nanoscale point contacts.
 
To study the conductance changes caused by ``jumps'' in the structural
configurations we relate them to the changes 
of the interference contribution to the backscattered electron flux 
in the contact region. The weak scattering approximation allows us to
neglect multiple scattering, restricting our calculations to 
a minimal number of scattering events. This corresponds to
the concept of local interference which
has been introduced in Ref. \onlinecite{Pelz} for the case of scattering by
complex defects \cite{Martin}. Mesoscopic effects in bulk samples caused by 
local interference of a more general type have been considered in Ref.
\onlinecite{18}. 

The standard approach to the conductance $G$ of small contacts
exploits the Landauer formula (see e.g. Ref. \onlinecite{Landauer})
\begin{equation}\label{landauer}
G = \frac{e^2}{h}\sum_{\alpha}(1 - R_{\alpha})
\end{equation}
where $R_{\alpha}$ is the reflection coefficient for
channel (mode) $\alpha$. In a simulation one needs to 
identify these modes $\alpha$.  
This can easily be done for
2D quantum point contacts with smooth ``adiabatic" boundaries
\cite{Glazman} whereas for realistic structures one 
meets with considerable problems. In particular we mention
Ref. \onlinecite{Kirczenow} where for the case of a short 2D
channel the conductance is calculated numerically by 
matching the ``bulk" plane wave modes to the channel's
transversal modes.

Here, we will apply the ``wave optics" approach which was 
developed for the transport in structures
with a small number of scattering events 
and is in the spirit of Ref. \onlinecite{Kirczenow}.
In Ref. \onlinecite{kozub} this approach was developed and successfully 
applied to analyze experimental data on mesoscopic transport in 3D ballistic  
point contacts with diffusive surroundings.
We also want to mention
the simulations of 
quasi-ballistic transport in static disordered structures \cite{Anna}
based on the Green's function method whose results  
agree well with those of Ref. \onlinecite{kozub}.

The ``wave optics" approach exploits the fact that for metallic point contacts
transport is always semiclassical in the sense that
the number of quantum channels is large, corresponding to the 
``geometrical optics" approximation for the de Broglie waves.
For an ideal point contact, separating the bulk leads, 
the modes $\alpha$ correspond to the ``optical
beams" cut by the orifice from the incident plane waves
$\bf k$. The
``quantization" of the modes originates from the Fourier expansion of the
kernel of the Fresnel--Kirchhoff integral which  
describes the diffraction of the
incident plane wave by the aperture in an opaque screen, 
see e.g. Ref. \onlinecite{Born}.
The number of the modes, $\sim (k_Fa)^2$  
(with ${\bf k}_F$ the Fermi wave vector), is the number of independent
solutions of the diffraction problem for a given
$|\bf k|\approx k_F$.

To calculate the backscattering coefficients $R_{\alpha}$
within this approach
we use the standard perturbation theory of scattering with  
Schr\"odinger's equation \cite{Landau}. We consider the
scattering of the incoming modes $\psi_{\alpha}$ which, as discussed above, 
at distances
of the order of the atomic scale can be approximated by 
plane waves
with wave vectors ${\bf k}_{\alpha}$. 
The resulting wave, scattered by a scatterer
$i$
at point ${\bf R}_i$, is
\begin{eqnarray} \label{eq:2}
\psi_{s,\alpha}(i,{\bf r}) = { \psi_{\alpha}({\bf R}_i)F_i
\over \mid {\bf r} - {\bf R}_i \mid } \cdot   
\exp \left( ik \mid {\bf r} - {\bf R}_i \mid)\right ) .  
\end{eqnarray}
Here $F_{i} $ is the scattering amplitude which in the following we assume  
to be equal for
all scatterers ($F = (s /4\pi)^{1/2}$).  
Including multiple scattering would 
add higher order terms in $F/a$. For a random 
distribution of 
scatterers the sum over all possible configurations would 
give a factor $\sim N^{1/2}$ to the correction of the wave function Eq.
\ref{eq:2}. 
Thus the total contribution of trajectories involving multiple scattering 
is by a factor $s^{1/2}N^{1/2}/(a(4\pi)^{1/2}) = s^{1/2}
n_i^{1/2}a^{1/2}/(4\pi)^{1/2} << 1$ smaller than
the contribution due to single scattering, i.e. neglecting multiple 
scattering is justified.

The coefficient $R_{\alpha}$ is by definition the ratio of
the ``backscattered" current 
to the ``incident" current of the mode $\alpha$.    
In our approximation we have:
\begin{equation}\label{curr}
R_{\alpha} = \frac{1}{v_{\perp,\alpha}|\psi_{\alpha}|^2} 
\int_S {\rm d}^2 {\bf r} {ie \hbar \over 2m}(<\psi_{s,\alpha}^{\ast}| \nabla
|\psi_{s,\alpha}> +
c.c.) .
\end{equation}
Here $\psi_{s,\alpha}$ is the sum of the
waves scattered by the scatterers
$i$, $v_{\perp,\alpha}$ is the velocity of the incident electron wave 
perpendicular
to the aperture. 
The apparent divergency due to $v_{\perp,\alpha}$ in Eq. 
\ref{curr} is integrable due to $v_{\perp,\alpha} \propto (k_F -
k_{\alpha,\perp})^{1/2}$, with ${\bf k}_{\alpha,\perp}$ the
projection of ${\bf k}_{\alpha}$ onto the orifice plane,  
and does not affect the final result
of Eq. \ref{eq:3}. 
$S$ is the orifice area while
$\bf r$ is the 2D radius-vector in the detector plane. 
Fig.\ \ref{scatter}  
depicts schematically two backscattered trajectories which interfere in the
detector plane.

From Eqs. \ref{eq:2} and \ref{curr} we get for the change of 
the conductance due to backscattering:
\begin{eqnarray} \label{eq:3} 
W_b \equiv \frac{\delta G_b}{G} =
{F^2 \over  \sum_{\alpha}(1-R_{\alpha})} \cdot
\sum_{\alpha}\int_S {\rm d}^2{\bf r}  \cdot   
\nonumber
\      \\ 
\sum_{i,j} {\cos \left( k_F(\mid {\bf r -
R}_i \mid - \mid {\bf r - R}_j \mid) + {\bf k}_{\alpha}
({\bf R}_i - {\bf R}_j) \right)
\over \mid {\bf r - R}_i \mid \mid {\bf r - R}_j \mid } \cdot
\nonumber
\     \\
( \cos (\angle (OX, {\bf r - R}_i)) + \cos (\angle (OX, {\bf r - R}_j))) 
\hskip 2cm 
\end{eqnarray}
Here OX is the contact axis. 
Eq. \ref{eq:3} is the basic input for the numerical studies described below.
The advantage of the coordinate representation used in our approach
is its direct contact to the simulation of electron 
transport through the structure.
In our calculations we consider amorphous structures 
where all atoms act as scatterers. For relatively
clean crystalline structures only defect atoms are responsible
for backscattering. We are mainly interested in the changes of the 
backscattering
due to changes of the structures and use all atomic positions as input 
for Eq.\ \ref{eq:3}. 
In the absence of
defects one would obtain the conductance of purely ballistic contact 
from the Landauer formula of Eq. \ref{landauer}. 

\section{Computational details and results}
With the help of Eq. \ref{eq:3} we carried out
numerical studies of the backscattered 
current for amorphous and crystalline monatomic metallic
structures and for amorphous selenium. 
We would like to note, however, that dealing with monatomic 
metallic glasses one meets difficulties
due to their strong tendency to crystallize.
This problem is known both from experiments on real metallic glasses
and from computer simulations. One 
consequence is that the structural relaxations
are mostly irreversible, preventing the formation
of multistable structures supposed to give rise to the
``fluctuators". We will discuss these problems in more detail
in relation to the results of our simulations.
Additionally we have calculated the backscattered electron flux by
using a more complex structure 
of amorphous Se, which is
much more stable against crystallisation. 
The results of the  
MD-simulation \cite{OS93,OS95} have 
shown that both soft sphere and
Se glasses 
behave similarly with respect to their dynamics which in both 
systems at low temperature exhibits local
relaxations. Although electron transport in real Se-based structures  
is hardly similar to that in  
metals, we believe that the simulations 
for the corresponding model
structures are in any case instructive. Indeed, one should
keep in mind that ``fluctuators" are thought to be 
related to defect regions where the structure
might be even more complex than in a monatomic metallic
glass.

The metallic structures 
were generated in earlier MD--simulations \cite{LS,SOL,OS95}. 
The interatomic interaction in these systems is given by a 
soft sphere potential $V(r) = \epsilon(\sigma/r)^{6} + A(r/\sigma)^4 + B$
where the energy-scale is given in units of $\epsilon$ and the length-scale 
in units of $\sigma$. In the simulations we used the units
$\epsilon = \sigma = m = 1$, $m$ being the atomic mass.
The density is kept constant at $\rho\sigma^3 = 1$ and the nearest 
neighbour distance corresponds to about $1.1\sigma$.
To reduce the 
numerical effort the potential is cut off at $r_c/\sigma = 3.0$ and 
shifted by a  polynomial $A(r/\sigma)^4 + B$ with $A = 2.54\cdot
10^{-5}\epsilon$ and $B = -3.43\cdot 10^{-3}\epsilon$. Both the potential 
and the force are zero at the cut-off.   
Such a potential is appropriate to model metallic systems  
\cite{12}.  

As described by Laird and Schober \cite{LS} to bypass crystallization  
the model glassy structures 
were produced from well equilibrated liquids by a rapid quench. 
With  typical quench-rates of the order 
$0.25k_B/\sqrt{m\sigma^2/\epsilon}$ the configurations are cooled to roughly 
half the glass temperature $T_g$ and left there for more than 1200 
molecular dynamics time-steps (MDS), i.e. about 60 typical vibration
periods, to stabilize the potential energy.  
The local minima of the structures were found by minimizing the potential 
energy with a steepest descent/ conjugate gradient algorithm \cite{Fletcher}. 
To have a reasonable statistics for the structures consisting of 500 and 
1024 atoms in all 60 and 21 model glassy structures were produced, 
respectively. 

The structural changes of these metallic glasses were observed in a previous 
molecular-dynamics simulation \cite{SOL}. 
To study the dynamics of these soft sphere glasses they were heated
with help of the
``Velocity--Verlet''--algorithm
in stages  
to temperatures ranging  
from $0.05$ to $0.125~T_g$.  The molecular dynamics was done
with periodic boundary conditions 
to reduce surface effects. After heating the glasses to the desired 
temperature, the temperature was kept constant by scaling the velocities of 
the atoms. 
At each temperature the dynamics of the systems was followed for 9000 
time-steps, which corresponds roughly to 500 periods of a typical 
vibration. 

To observe more complex structures and to reduce further the ``finite size''
effects additionally 15 soft-sphere-glasses with $N=5488$ were generated 
from hot melts and
used to study relaxations in a greater detail \cite{OS95}. 
These glasses were heated to temperatures ranging from 0.05 up to 0.15~$T_g$ 
and left there for 90000~MDS (an order of magnitude longer than the smaller 
systems).   

During the molecular dynamics simulations at constant temperature 
the configurations move to new minima of the energy landscape   
due to local hopping processes over small energy-barriers. To detect new
configurations on the energy hypersurface the displacements of the atoms from
the metastable equilibrium positions ${\bf R}_n^i$ of atom $n$ in 
configuration $i$ are measured and the total displacement 
of the structure is defined as:
\begin{equation}
\Delta R^2(t) = \sum_n({\bf R}_n(t) - {\bf R}_n^i)^2
\end{equation} 
where ${\bf R}_n(t)$ is the actual position vector of atom $n$. 
If the total displacement of the atoms exceeds a cut-off value and the 
residence time of the atoms in the new positions also exceeds a minimal period
of at least three soft, low-frequency vibrations (several 100 MDS) the new 
positions
of the particles are accepted as a new minimum configuration. The cut-offs of
displacement and resident time, respectively, are chosen such that spurious 
minima are avoided. 
This procedure is described in greater detail in Ref. \onlinecite{SOL,OS95}. 

These minima are monitored during the molecular dynamics runs. 
The corresponding relaxations from one minimum to another can be either
reversible or irreversible. 
The irreversible jumps are mainly due
to the instability of the monatomic metallic glasses. 
This instability manifests itself in the fact that 
during the observation time some of the samples partially crystallized.
Such a crystallisation is connected with a global rearrangement of nearly all 
atoms in the configuration. 
This effect depends strongly on the size of the samples used in the simulation.
The larger the system, the more stable are the configurations.
The largest systems show greater stability towards crystallization; just one
configuration (out of 15 with $N = 5488$ atoms) crystallized during the 
observation time at a temperature $T=0.15T_g$. 
But the tendency to lower the potential energy prevails during the simulation.
 
Therefore, the first jumps after heating up to different temperatures are
related to non-local rearrangements of the glasses, i.e. jumps with large 
displacements involving many atoms and leading to more stable configurations. 
These jumps are followed by relaxations located on only a few atoms. 
A measure for the localization of a jump is the effective mass 

\begin{equation} \label{eq:5}
M_{\rm eff} = m\cdot\frac{(\Delta R)^2}{|\Delta {\bf R}^2_{\rm max}|}
\end {equation}

where $(\Delta R)^2=\sum_n({\bf R}_n^i-{\bf R}_n^f)^2$
is the square of the total distance $\Delta R$ between two successive minimum
configurations (called ``initial'' and ``final'' positions of the jump).
${\bf R}_n^{i,f}$ denotes the respective 
initial and final positions of atom $n$, 
$|\Delta {\bf R}^2_{\rm max}|$ is the maximal distance a 
single atom jumps in this relaxation.
Since we take the masses of the atoms comprising 
the structures as equal, $M_{\rm eff}$ counts the effective number of particles 
contributing to the hopping process. 

In an extended computer simulation it is found that the displacement of
a single atom in the relaxation is only a fraction of the nearest neighbour 
distance (typically a tenth of a bond length), i.e. relaxations at low 
temperatures are local processes with small displacements of the atoms 
contributing to this thermally activated jump. 
However, even the small displacements of the single activated atoms exceed 
the vibration amplitudes. 
Since the typical total 
distance between two minimum configurations is about one or several nearest 
neighbour distances \cite{Olig-Schober}, the effective mass is of the order
of 10 to 100 atoms contributing essentially to the relaxations. 
It is found that these entities of some ten particles involved in the 
hopping processes form  chain-like structures which move collectively 
along these chains \cite{Schober97}. 

Since we are interested in 
multistable configurations, we restrict ourselves to the
relatively rare reversible changes of the configurations, which means that 
the initial
configuration and the final one of successive jumps are identical.  
A measure for the reversibility of successive jumps
can be given by the quantity $c_{RR'}$

\begin{equation} \label{eq:4}
c_{RR'} = \frac{1}{\Delta R\Delta R'} 
\sum_{n=1}^N |\Delta{\bf R}_n| |\Delta{\bf R}'_n| 
\end {equation}

where $\Delta {\bf R}_n=({\bf R}_n^i-{\bf R}_n^f$) is the displacement vector
of atom $n$ in the relaxation $\Delta R$
(${\bf R}_n^{i,f}$ being, as above,     
the initial and the final positions of atom $n$ in the relaxation $\Delta R$), 
and $\Delta R$ and $\Delta R'$ are the total jump-lengths of two 
successive relaxations
given by  $\Delta R^2 = \sum_{n=1}^{N}(\Delta{\bf R}_n^2)$.
For a reversible jump one has $c_{RR'} = 1$, i.e. 
the same atoms are contributing equally to both jumps. 
If the jumps are uncorrelated, $c_{RR'}$ would be of the order of $c_{RR'}
= N_{p}/N$, where $N_{p}$ is the number of atoms involved in the hopping. 
During the MD-runs of the 60 glasses with 500 atoms only two structures 
undergo reversible jumps, whereas for the glasses with 1024 soft spheres
only one reversible relaxation occurred. The effective masses of the  
reversible relaxations observed in soft-sphere-glasses with $N=500$ and $1024$ 
atoms range from  6 to 16 atomic masses. 
At $T \approx 0.1 T_g$ half of the glasses with $N = 5488$ settled during the 
observation time to configurations where reversible jumps involving up to four 
local minima are observed.  
For the simulations with 5488 atoms reversible relaxations with effective 
masses from 9 to 50 atoms were found. 
      
In order to calculate 
the electron back-flow we use incident electron waves with
{\bf k}-vectors of length $k=\pi/\sigma$. 
This value determines the ``discretization'' 
of the numerical integration in Eq. \ref{eq:3}.
The summation over the modes in Eq. \ref{eq:3} 
becomes a sum over discrete angles which we take as
$\Delta \phi = 2\pi /{k a}$ and $\Delta \theta = \pi /{k a}$, 
where $a$ is the
side-length of the amorphous sample, for the metallic glasses the
side-length is $a= N^{1/3} \sigma$.
Using different grids for the numerical discretization 
changes the results by a few percent only, 
i.e. they are not crucially sensitive to the 
specification of the modes.
To escape numerical problems with vanishing denominators
in Eq. \ref{eq:3} we integrated over $r$ not in the
front orifice plane but instead in a ``detector" plane
shifted by  
approximately one atomic distance from the orifice plane. 
The integration becomes a double-sum with $x = i\Delta x$ and $y=j\Delta y$  
where $\Delta x = \Delta y = 1/k$.
The detector plane is taken about 2 to 3 times larger than the orifice; 
the thus neglected backscattered current  
 is less than $10-25 \%$. 
As value of the scattering amplitude $F$,
we take (typical for metals) $k_F F \simeq 1$. 
One can show that for a relatively small contact volume (small $a$ and $N$) 
the weak scattering approximation 
($s N /a^2 < 1$) holds with this assumption. 

To test our procedure we applied it to the simple case of a self 
interstitial atom in an fcc metal. This so called dumbbell configuration is
formed by replacing one atom by a pair of atoms aligned in (100) direction 
\cite{ERS86}. The elementary jump process of this self interstitial atom
configuration is to a nearest neighbour site and involves a change of 
orientation to a different cubic axis, i.e. the dumbbell is now aligned in
(010) or (001) direction. 
Using the soft sphere potential we construct configurations comprising 501 
atoms with the 
dumbbell in the middle of the structure aligned in (100) and (001) directions.
The coordinates of the 
atoms are then used as input for Eq. \ref{eq:3} to calculate the 
backscattering of these structures. 
For the configuration with the dumbbell in (100) direction, i.e. perpendicular 
to the detector plane (see Fig.\ \ref{scatter}) we find $W_b = 0.536$, and for 
the two other orientations (001) and equivalently the (010) alignment, 
parallel to the detector plane, we calculate $W_b = 0.540$.
From this we deduce that the relative magnitude of the fluctuation due to the
reorientation of the dumbbell is about $0.7\%$.

In the following we present the relative magnitudes of the
fluctuations caused by jumps and relaxations in monatomic metallic glasses
and defect structures as described above. 
Since we are interested in the changes of the backscattered flow due
to relaxations of the samples we determine the minima observed in the  
MD runs and calculate $W_b$ for these structures corresponding to the local
minima of the potential energy. We neglect the contribution  of 
atomic vibrations to the electronic backscattering which will be similar for
different configurations. 
In Fig.\ \ref{cr6sb4} we show $W_b$ for a metallic glass with $N=500$ atoms.
After the initial rearrangement connected with a decrease in potential energy,
the system was stable for more than 20000 MDS before 
a reversible jump occurred with a
total jump-length of $\Delta R = 1.03\sigma $ and an effective mass 
$M_{\rm eff} = 16m$. 
The relative magnitude of the backscattered current is about $0.6\%$.

Fig.\ \ref{cr6gl21} shows $W_b$ for a system with $N=1024$ 
atoms.
After approximately 6000 time-steps a reversible jump occurred with a
total jump-length of $\Delta R = 0.44\sigma$ and an effective mass 
$M_{\rm eff} = 6m$. 
The fluctuation due to the structural relaxation is approximately
$0.2\%$ of the total interference contribution.

To simulate more complex multistable atomic aggregates 
larger samples are needed.  
Indeed as we have discussed the smaller systems are relatively unstable thus
preventing 
extended multistable atomic aggregates.  
Therefore, we have also studied
the current transport based on the results for a glass with $N = 5488$
atoms. To shorten simulations of the interference pattern we 
restricted the summations over the scattering atoms $R$ in Eq.\ \ref{eq:3} 
to a subset $N_s\approx 700$ of the atoms including and surrounding the 
active ones. 

The conductance change of this glass heated to $0.1 T_g$ and aged for 90000 MDS
is shown in 
Fig.\ \ref{cr6xl25}. In Fig.\ \ref{cr6xl25}(a) we show the signal of the
backscattering for the detector placed perpendicular to the arbitrarily 
chosen x-axis of the
structure, in Fig.\ \ref{cr6xl25}(b) we measure the
backscattered current in the plane perpendicular to the z-axis of the
point contact, i.e. we have
placed the detector parallel to the x-y-plane of the sample of the
configuration and the bulk leads are now perpendicular to the x-y-plane of
the nanoscale point contact.
The results of the backscattering are much more complex than the 
ones described above for the smaller systems:  
the complex behaviour of the conductance change shown in Fig.\ \ref{cr6xl25} 
results from 15 jumps between 
eight minima denoted by A,B,C,...,H according to their occurrence during the
simulation.                           
After an initial drop in energy the glass relaxes to a (meta)stable region in 
the configurational space, where we monitored these strongly correlated jumps 
between 8  minima of nearly equal energy. 
After about 17000 MDS a 
reversible jump (E$\rightarrow$F$\rightarrow$E) occurred with a total 
jump-length $\Delta R = 0.72\sigma$ and an effective mass $M_{\rm eff} = 16m$. 
At $t \approx 42000$ MDS the glass returns to minimum A 
and relaxes to minimum C ($\Delta R = 1.22\sigma$ and
$M_{\rm eff} = 18m$) and again back to minimum A. After visiting two other minima
G and H, the jump pattern (A$\rightarrow$C$\rightarrow$A) is repeated. 

The magnitude of the relative backscattered electron flux is about 
0.037$\%$ for the contribution of the first reversible relaxation 
(E$\rightarrow$F$\rightarrow$ E) measured in a 
detector plane perpendicular to the x-axis  of the sample (see the experimental
 setup in Fig.\ \ref{scatter}) 
and 0.032$\%$ for the second pronounced
fluctuation (A$\rightarrow$C$\rightarrow$A).

The signal of these fluctuators changes drastically when we measure the
backscattered current in the plane perpendicular to the z-axis of the 
configuration. As described above we have
placed the detector parallel to the x-y-plane of the sample. The change of 
the experimental setup enables us to detect inhomogeneities in the structure
and in the dynamics of the nanoscale point contact. 
The first reversible jump (E$\rightarrow$F$\rightarrow$ E) causes a change of 
the relative magnitude of the   
backscattered electron flux  
of approximately 0.11$\%$ and the second one (A$\rightarrow$C$\rightarrow$A) 
leads to a relative change
in the current of about 0.66$\%$.     
These changes of the signals with respect to the axis may be due to the strong
displacements of the atoms in 
z-direction in both reversible relaxations. 

Additionally we calculated the 
backscattered current also for a partially crystallized 
defect structure of $N=500$ atoms. After heating to $0.05T_g$ 
the configuration begins to crystallize. During the observation time the
sample relaxes to another defect configuration. The initial relaxation is 
accompanied
by a large drop in energy ($\Delta E = 30.28\epsilon$) while the following 
jumps lead to minima with similar  
energies ($\Delta E <  10^{-7}\epsilon$), in the last relaxation the system
jumps to a minimum with slightly higher energy than the one preceding  
($\Delta E = 0.00670\epsilon$).  
The corresponding jump-length is $\Delta R = 13.70\sigma$ for the initial 
relaxation.   
For the three following jumps ($\Delta R = 5.23\sigma,
7.75\sigma, 6.83\sigma$) with small energy changes we find effective 
masses 
$M_{\rm eff} = 68m, 74m, 72m$, respectively. For the last jump
the effective mass is about $74m$ and the jump-length is about $6.53\sigma$. 
The jumps following the initial decrease in energy are strongly correlated, 
i.e. mostly   
the same atoms are involved. Fig.\ \ref{cr6g58} shows 
the results for
the calculations of the backscattering flow $W_b$ plotted vs. the observation 
time.
The amplitude of the fluctuations after the initial step is about $3\%$.
   
So far we have considered monatomic metallic glasses, where due to
crystallization effects long observation times are not accessible.  
Compared to the soft sphere glasses the models for selenium are
more stable, and longer observation times are possible without
an appreciable drop in potential energy.
Therefore, we also report results of the backscattered electron 
flux, normalized with respects to the incident flow, for a 
selenium glass. Although selenium is not a metal, the
complex structure comprising chains and rings may be a good model for 
atomic entities forming the ``fluctuators".  

The interaction of the Se-atoms
is described by a short-range three-body-potential \cite{OJRS}.  
The glasses are produced by quenching 
corresponding liquids with quench-rates 
of $\dot T = 10^{12}-10^{14}$ K/s 
and pressures up to 10 GPa \cite{OS93}. Note that if the quenching
would be
too rapid density fluctuations of the liquid could be frozen in.
One 
possible way to avoid the formation of resulting micropores is to apply lower 
quench-rates and/or pressure. The glasses are heated to temperatures of about
$0.30T_g$ and then observed for approximately $0.5$ ns 
at every temperature (this corresponds to roughly 
3400 high frequency
vibrations).   
In Fig.\ \ref{seg39} we plot the backscattered flow $W_b$ versus time for 
a glass 
of $N = 750$ atoms at $T \approx 0.04 T_g$. One can clearly observe a
telegraph signal caused by a reversible jump occurring four times during 
the observation period. Its effective
mass is $M_{\rm eff} = 10$ and its total jump-length
$\Delta R = 0.5$\AA. The amplitude of the fluctuation is about $0.42\%$ of 
the total
interference contribution to the backflow.

\section{Discussion and conclusions}

We calculate the conductance fluctuations via the standard Landauer formula
from the change of backscattering by the atoms comprising the junction.
For the small junctions considered here electron scattering can be treated  
ballistically in the weak scattering limit. 

In geometries where multiple scattering is important different methods have 
to be used. Todorov and Sutton, and Bratkovsky et al. combine a molecular
dynamics simulation with a T-matrix approach in tight binding approximation 
to calculate the conductance through atomic-scale metallic contacts during
pulloff \cite{Todorov,Bratkovsky,Todorov96}. In agreement with experiment they 
observe conductance jumps due to atomic rearrangements in the tip. 

We want to point out that our approach is not suitable to reproduce the   
limiting case where the point contact merges into the crystalline structure.
In this limit the conductance is no longer dominated by the scattering of the 
atoms in the junctions and the aperture.

Let us now consider the characteristic magnitude of the non-stationary
current fluctuations. As seen from Eq. \ref{eq:3}, one can discriminate between
two terms: (i) 
``classical" backscattering related to the diagonal terms in the sum
of Eq. \ref{eq:3} which is independent of
the coordinates of the scatterers and thus not sensitive to 
relaxations; and (ii) 
the interference contribution due to the $N(N-1)$ non-diagonal 
terms proportional to cosines and depending on the positions of the atoms.
For a given mode $\bf k$ and for
a given position $\bf r$ in the
detector plane the standard deviation of the
interference term is $\sim \sqrt {N(N-1)} \sim N$. However, additional
averaging over the $\sim N^{2/3}$ different modes 
 and over the different
positions of $\bf r$ 
reduces the standard deviation 
to $N/(N^{2/3})^{1/2}(N^{2/3})^{1/2} = N^{1/3}$. 
Here positions separated by distances larger
than $\sim 1/k$ are to be considered as independent, concerning the
interference, giving in our units about
$N^{2/3}$ independent positions.
Therefore, the total  
standard deviation of the interference contribution due to a
displacement of all scatterers is smaller 
than the average backscattered flow by about a factor of $N/N^{1/3} = N^{2/3}$. 
If one    
considers the current transport according to the concept of
``quantum channels", where in the weak scattering limit each 
elementary cell in the point contact orifice 
roughly corresponds to one channel,
the standard deviation of the interference contribution 
is of the order of the contribution of one channel, 
in agreement with the picture of Universal Conductance
Fluctuations (see e.g. Ref. \onlinecite{Feng}).

For a shift of a single scatterer (affecting $N$ terms in the summation
over $i,j$ in Eq. \ref{eq:3})
the effect is expected to be reduced by   
$N^{1/2}$ while for an uncorrelated jump of $M$ atoms
the expected effect is $\sim M^{1/2}/N^{2/3}N^{1/2} \sim
M^{1/2}N^{-7/6}$ (normalized with respect to the average backflow
current). Thus a jump of about $15$ atoms out of $500$ should
give a fluctuation of relative magnitude
$\sim 0.003$. 

As seen, the order of magnitude of the experimental results
is in agreement with this estimate. Some larger effects
shown in Fig.\ \ref{cr6g58}
 can be related to the fact that
actually the jumps cannot be considered as incoherent shifts
of $M$ atoms; the coherency of the atomic
motion can increase the observed effect.
Qualitatively, the results show the decrease of the relative fluctuation 
with increasing $N$ and with decreasing $M$.
The magnitude of the
fluctuations gives information about the size of
the relaxing region and about the character of the motion.

The occurrence rate of the fluctuations is 
many orders of magnitude smaller than characteristic vibrational
frequencies of the structure and thus we are obviously dealing  
with slowly relaxing degrees of freedom. However, for the
simulated structures the rate is still much larger than a typical frequency  
for the telegraph noise observed in real experiments  
\cite{1,2}. The reason is again the instability of
the structures studied with respect to crystallization
which simply prevents long observation times which is
in particular related to the small sizes of the systems and to
the role of the periodic boundary conditions.   
In real systems relaxations would be slowed down due to strains
exerted by the embedding material. 
In our point of view the simulations give a good model
for the slowly relaxing multistable entities in real
structures.
The structures in real point contacts
will be somewhere in between the amorphous and partially crystalline state. 
In this respect we would especially like to emphasize that 
the multistable defects
were also demonstrated for the simulated  crystalline
defect structure of Fig.\ \ref{cr6g58}.

The simulated current fluctuations exhibit a rather complex time behaviour.
The complex behaviour shown in Fig.\ \ref{cr6xl25} results from jumps connecting 
eight different configurations of nearly equal energy. Since the initial and 
final configurations are identical all jumps are reversible. Some jumps might
trigger each other, others will occur randomly. E.g. after 42000 MDS a jump
followed immediately by the return jumps occurs, 15000 MDS later the
same jump is followed by two different jumps before returning to the start
configuration. 
Actually the simulated picture   
resembles the one observed for real metallic
contacts \cite{2} at high enough temperatures
referred to as ``fluctuator melting". We would like
to emphasize that the slowing down of relaxations with temperature prevented 
us from studies of the extremely low temperatures 
where simpler two state telegraph signals are expected.   
We want to emphasize, however, that the
temperatures of our simulations are still much
below  
the glass-transition and melting temperatures. We see
a similarity to real experiments \cite{2} where 
``fluctuator melting" was observed at temperatures of about $100$ K, 
i.e. much below the melting temperature of the metal.

Our simulations show that typically  
the ``fluctuators" involve large groups of atoms 
comprising up to 100 particles 
arranged mainly along some preferential direction. In Fig.\ \ref{r6_relax}  
we show such a fluctuator identified by the of MD-simulations for
the current noise shown in Fig.\ \ref{cr6xl25}. The atoms contributing 
strongly 
to the relaxation are shown with their respective initial and final positions 
marked as full and shaded circles. The structure formed by the 
particles resembles
a chain with side branches. Note that the displacement of the atoms from 
the initial
minimum to the second one is nearly one-directional (along the chain). 
This fluctuator corresponds in Fig.\ \ref{cr6xl25}  
to the pronounced reversible relaxation after 42000 MDS with an effective mass 
$M_{\rm eff}=18m$ and a total displacement $\Delta R=1.22\sigma$. 

Multistable configurations are more easily formed in 
the more complex Se structure. To some extent this hints towards 
a possible role of complex defects in the formation of fluctuators.
In Fig.\ \ref{se39rel} we show the atoms, which strongly participate 
in the telegraph noise shown in Fig.\ \ref{seg39} at $t < 250$ ps.          
The complexity of the relaxation is
reflected by the fact that atoms of two regions of the system 
strongly participate in the jump. One of the structures of the relaxing
entity is nearly one-dimensional
and the atoms comprising this structure move cooperatively
along this preferential direction.  
In particular, one can speculate on the apparent similarity between the helical 
structure in Se and the structure near the core of a screw
dislocation. In this respect we would like to mention the
experimental studies \cite{disloc} where the observed $1/f$ noise 
was shown to originate from atomic motion near dislocation
cores.

One notes that the picture of the current behaviour is more
complex than simple fluctuations which average out over long
enough time scales which are not accessible in MD-simulations. 
We observe a number of relaxations and related current fluctuations
with different relaxation times. The time window of our simulation 
is too narrow to determine whether these fluctuations obey a $1/f$ 
law in the large ensemble or long time limits.

To conclude, 
in the framework of a simple model   
we have done a first simulation of multistable current noise
in disordered contacts. The slowly relaxing modes responsible for the
noise are visualized. The dependence of the noise magnitude on the size
and structure of the modes was studied. In particular, the role of
coherent atomic motion was emphasized. 
An important feature is a drastic change in the fluctuator dynamics
(a transition to a complex multistable character of fluctuations)
at temperatures of about a tenth of 
the melting temperature of the sample;
this behaviour agrees with ``fluctuators melting" reported  
for real metallic point contacts \cite{2}. 

\acknowledgments 
We are indebted to H.R. Schober and V.L. Gurevich for reading the manuscript 
and for many valuable remarks and discussions. One of us (V.I.K.) acknowledges
the hospitality of the Institut f\"ur  Festk\"orperforschung of
the Forschungszentrum J\"ulich and the financial support by the
German Ministry of Technology under the Russian-German scientific
cooperation agreement. One of us (C.O.) gratefully acknowledges funding by 
the Deutsche Forschungsgemeinschaft
through the Sonderforschungsbereich 408.

\begin{figure}[h] 
\caption{Schematic plot of backscattered trajectories each involving 
one atom acting as scatterer of the incident electron wave.
The incident wave is represented by its wave vector ${\bf k}$.
Filled circles depict some of the atoms in the point contact.
$R_i$ and $R_j$ denote the  positions of the atoms and 
$r$ the two-dimensional radius vector in the detector plane.}  

\label{scatter}
\end{figure}   
\begin{figure}
\caption{Fluctuator behaviour of a configuration with $N = 500$ atoms.
$W_{b}$ is plotted against time in units of MDS. The reversible relaxation 
at $t \approx 25000$MDS has an effective mass $M_{\rm eff} = 16m$ and a 
total jump-length $\Delta R = 1.03\sigma$.}
\label{cr6sb4}
\end{figure}   
\begin{figure}
\caption{Fluctuator behaviour of a configuration with $N = 1024$ atoms.
$W_{b}$ is plotted against time in units of MDS. The reversible relaxation 
at $t \approx 6000$MDS has an effective mass $M_{\rm eff} = 6m$ and a 
total jump-length $\Delta R =0.44\sigma$.}
\label{cr6gl21}
\end{figure}

\begin{figure}
%

\caption{Fluctuator behaviour in a glass of 
5488 soft spheres undergoing fifteen jumps between eight minima denoted by
A,B,C,...,H. Different directions of   
detector planes are shown: (a) plane $\perp$ x-axis and 
(b) plane $\perp$ z-axis.
Two direct reversible relaxations (with effective masses
$M_{\rm eff} = 16m$ and $M_{\rm eff} = 18m$ and corresponding jump-lengths 
$\Delta R = 0.72\sigma$
and $\Delta R = 1.22\sigma$) can be observed.
$W_{b}$ is plotted against time in units of MDS.}
\label{cr6xl25}
\end{figure}

\begin{figure}
\caption{Fluctuator behaviour of a configuration with $N = 500$ atoms, which
partially crystallized at a temperature $T = 0.05T_g$. The last three jumps
with effective masses $M_{\rm eff} = 74m,72m,74m$ have jump-lengths 
$\Delta R = 7.75\sigma,
6.83\sigma$ and $6.53\sigma$, respectively. These relaxations are strongly
correlated with $c_{RR'} = 0.958, 0.996$.
$W_{b}$ is plotted against time in units of MDS.}
\label{cr6g58}
\end{figure}
\begin{figure}
\caption{Fluctuator behaviour of a configuration with $N = 750$ Se-atoms,
at a temperature $T = 0.04T_g$. During the time-range $t = 15-260$ ps
the glass switches between two configurations. The effective mass of the
reversible jumps is $M_{\rm eff} \approx 10$ and the jump-length $\Delta R \approx
0.5$\AA. The signal of the fluctuation is about $0.42\%$ of the total
contribution. $W_{b}$ is plotted against time in units of ps.}
\label{seg39}
\end{figure}
\begin{figure}
\caption{Reversible relaxation at $T = 0.1 T_g$ observed in a  
glass with $N = 5488$ atoms. Full spheres depict the initial positions 
of the atoms, shaded spheres are the atomic positions after the relaxation.  
Shown are all atoms with a displacement more   
than 0.4 of the maximal contribution to this relaxation. The effective mass
of this relaxation is $M_{\rm eff} = 18m$ and the total jump-length is 
$\Delta R = 1.22\sigma$. The largest contribution of a single atom to this jump
is $\Delta R_{\rm max} = 0.29\sigma$. 
The backscattering contribution of this
jump is shown in Fig.\ \ref{cr6xl25}.}
\label{r6_relax}
\end{figure}
\begin{figure}
\caption{Reversible relaxation at $T = 0.04 T_g$ observed in
a Se-glass with $N = 750$ atoms. Full spheres depict the initial positions
of the atoms, shaded spheres are the atomic positions after the relaxation.
Shown are all atoms with a displacement more
than 0.2 of the maximal contribution to this relaxation. The maximal 
contribution to this relaxation is $\Delta R_{\rm max} = 0.44$ \AA . 
The backscattering contribution of this
jump is shown in Fig.\ \ref{seg39}.}
\label{se39rel}
\end{figure}

\begin{thebibliography}{10}
\bibitem[*]{VIKaddr}Permanent address: A.F.Ioffe Physico-Technical Institute, 
194021 St.-Petersburg, Russia  
\bibitem[**]{COaddr}Present address: Institut f\"ur Physikalische und 
Theoretische Chemie, Universit\"at Bonn, D-53115 Bonn, and Institut 
f\"ur Algorithmen und
Wissenschaftliches Rechnen, GMD -- Forschungszentrum Informationstechnik,
D-53754 Sankt Augustin, Germany. 
\bibitem{1}
C.T. Rogers and R.A. Buhrman, Phys. Rev. Lett.
{\bf 53}, 1272 (1984).
\bibitem{2}
K.S. Ralls and R.A. Buhrman, Phys. Rev. Lett.
{\bf 60}, 2434 (1988).
\bibitem{3}
A.M. Kogan and K.E. Nagaev, Solid State Commun.
{\bf 49}, 387 (1984).
\bibitem{4}
A. Ludviksson, R. Kree and A. Schmid, Phys. Rev. Lett.
{\bf 52}, 950 (1984).
\bibitem{5}
V.I. Kozub, Sov. Phys. - JETP, {\bf 86}, 1303 (1984).
\bibitem{6}
V.I. Kozub, Sov. Phys. - Solid State, {\bf 26}, 1851 (1984).
\bibitem{7}
V.I. Kozub, Sov. Phys. - JETP, {\bf 60}, 818 (1984). 
\bibitem{8}
Yu.M. Galperin, V.G. Karpov and V.I. Kozub, Adv. in Phys. {\bf 38}, 669 
(1989).
\bibitem{Muller} C.J. Muller, J.M. van Ruitenbeck, and L.J. de Jongh,
Phys. Rev. Lett. {\bf 69}, 140 (1992).
\bibitem{9}
V.I. Kozub and A.M. Rudin, Phys. Rev. B {\bf 47}, 13737 (1993).
\bibitem{LS}
B.B. Laird and H.R. Schober, Phys. Rev. Lett. {\bf 66}, 636 (1991).
\   \\
H.R. Schober and B.B. Laird, Phys. Rev. B {\bf 44}, 6746 (1991).
\bibitem{SOL}
H.R. Schober, C. Oligschleger and B.B. Laird, J. Noncryst. Sol.
{\bf 156-158}, 965 (1993).
\ \\
H.R. Schober and C. Oligschleger, Phys. Rev. B {\bf 53}, 11469 (1996) 
\bibitem{Pelz} J. Pelz and J. Clarke, Phys. Rev. B {\bf 36}, 4479 (1987).
\bibitem{Martin}
J.W. Martin, Phil. Mag. {\bf 24}, 555 (1971).
\bibitem{18}
Yu.M. Galperin and V.I. Kozub, Sov. Phys. - JETP, {\bf 73}, 179 
(1991).
\bibitem{Landauer} C.W.J. Beenakker, H. van Houten,
Quantum transport in semiconductor nanostructures
in: Solid State Physics, {\bf 44}, Academic Press, 1991.
\bibitem{Glazman} L.I. Glazman et al., JETP Lett. {\bf 48}, 301 (1988)
\bibitem{Kirczenow} G. Kirczenow, Solid State Comm. {\bf 68}, 715
(1988).
\bibitem{kozub}
V.I. Kozub, J. Caro and P.A.M. Holweg, Phys. Rev. B {\bf 50}, 15126
(1994)
\bibitem{Anna} A. Grincwajg, G. Edwards and D.K. Ferry,
Physica B {\bf 227}, 54 (1996)
\bibitem{Born} M. Born and E. Wolf, {\it Principles of optics: Electromagnetic
Theory of Propagation and Diffraction of Light}, Pergamon Press,
London, 1959, pp.374. 
\bibitem{Landau} L.D.Landau, E.M.Lifshitz, {\it Quantum mechanics},Pergamon,
Oxford, 1972. 
\bibitem{OS93}
C. Oligschleger and H.R. Schober, Physica A 201, 391-394 (1993)
\bibitem{OS95}
C. Oligschleger and H.R. Schober, Solid State Comm. Vol.93, No. 12, 1031-1035
(1995)
\bibitem{12} W.G. Hoover, S.G. Gray, and K.W. Johnson, 
J. Chem. Phys. {\bf 55}, 1128 (1971).
\bibitem{Fletcher} R. Fletcher, and C.M. Reeves, Comput. J. {\bf 7}, 149
(1964).
\bibitem{Olig-Schober}
C. Oligschleger and H.R. Schober, submitted                                  
\bibitem{Schober97}
H.R. Schober, C. Gaukel, and C. Oligschleger, Defect and Diffusion Forum 
{\bf 143-147}, 723 (1997).
\bibitem{ERS86}P. Ehrhart, K.H. Robrock, and H.R. Schober, in {\it Physics of
Radiation effects in Crystals} (R.A. Jonson and A.N Orlov eds.), p.3,
North Holland, Amsterdam: 1986.  
\bibitem{OJRS}
C. Oligschleger, R.O. Jones, S.M. Reimann, and H.R. Schober, 
Phys. Rev. B {\bf 53}, 6165 (1996) 
\bibitem{Todorov} T.N. Todorov and A.P. Sutton,
Phys. Rev. Lett. {\bf 70}, 2138 (1993).
\bibitem{Bratkovsky} A.M. Bratkovsky, A.P. Sutton, and T.N. Todorov 
Phys. Rev. B {\bf 52}, 5036 (1995).
\bibitem{Todorov96} T.N. Todorov and A.P. Sutton,
Phys. Rev. B {\bf 54}, R14234 (1993).
\bibitem{Feng} S. Feng, P.A. Lee, and A.D. Stone, Phys. Rev. Lett. {\bf 56},  
1960 (1986)
\bibitem{disloc} E. Ochs, M.J.C. von Homberg, P.F.A. Alkemade, K. Armbruster,
A. Seeger, H. Stoll, and A.H. Verbruggen, in 
{\it Noise in Physical Systems and $1/f$ Fluctuations}: Proceedings of the 
14th International Conference (eds. C. Claeys and E. Simoen), p.415, 
World Scientific Pub. Co. (1997).
\end{thebibliography}
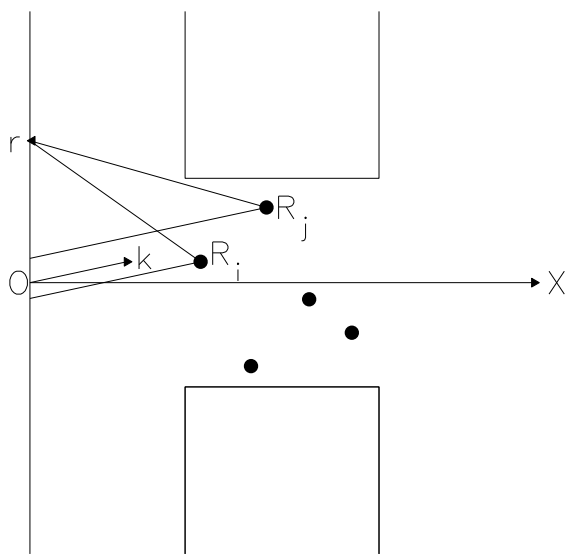
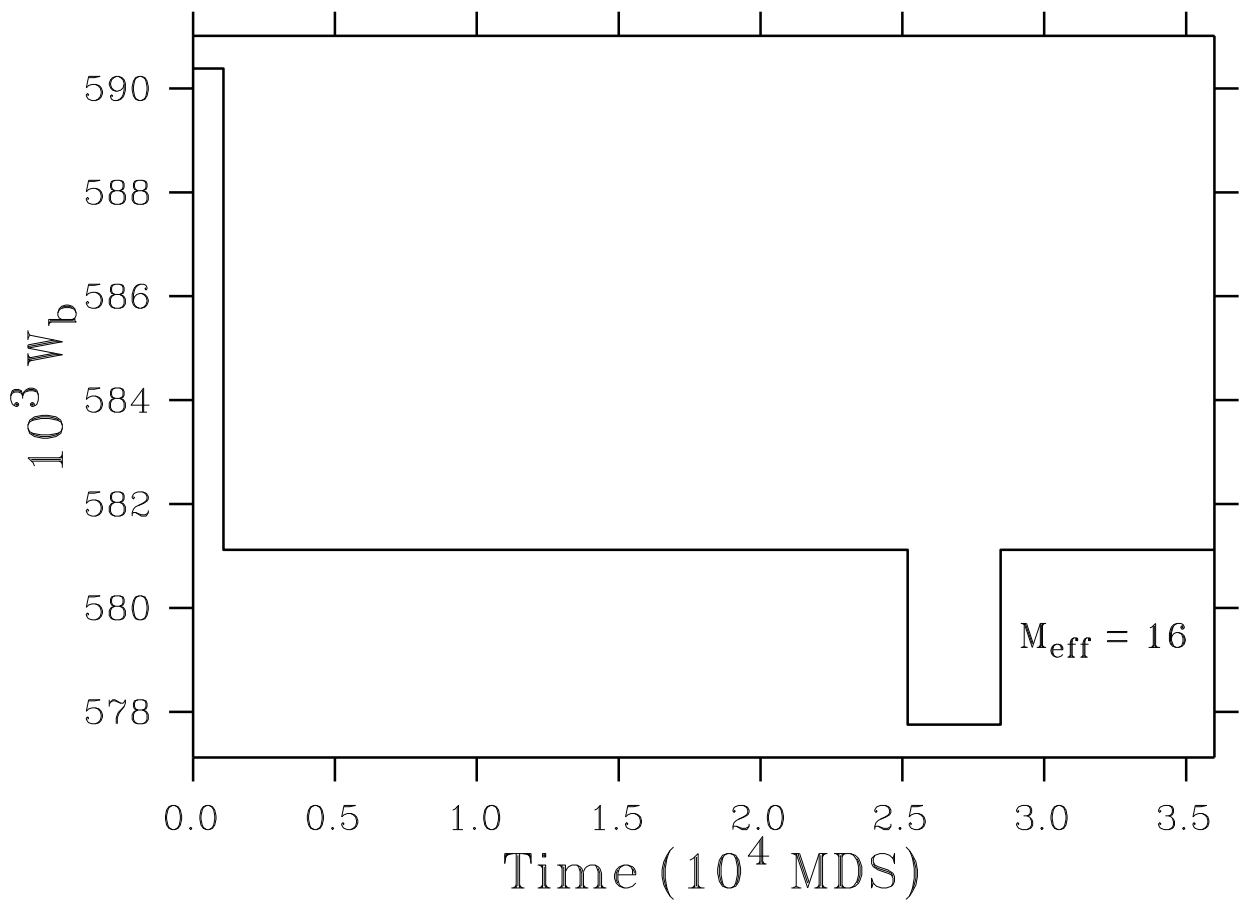
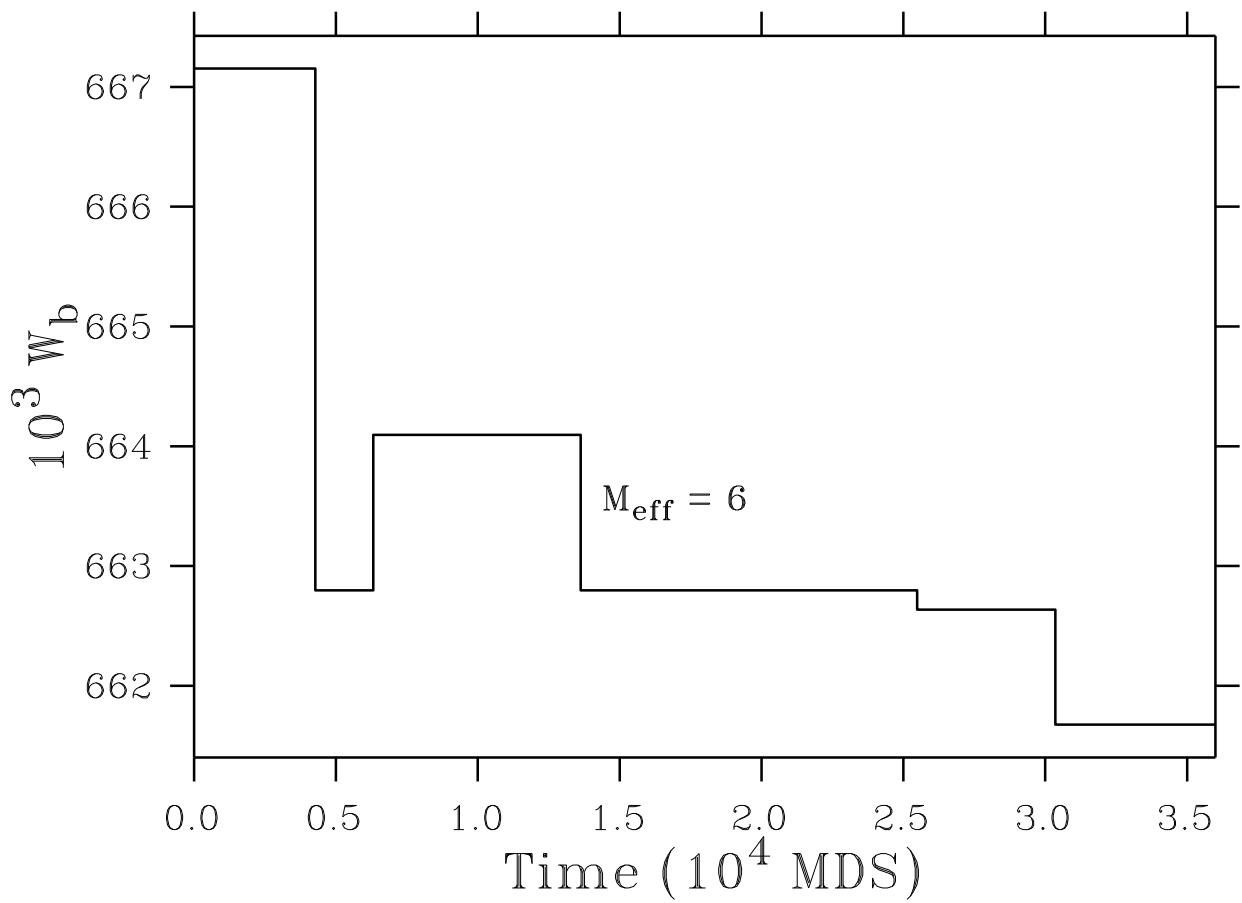
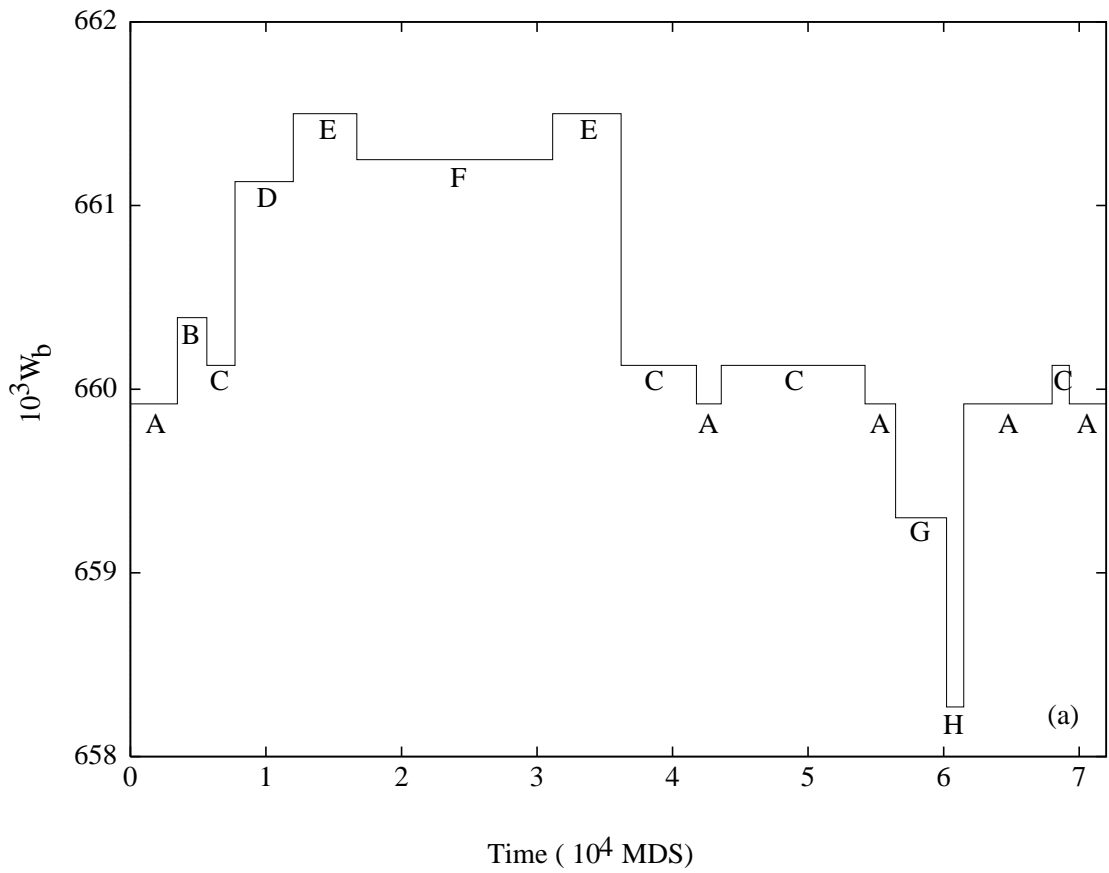
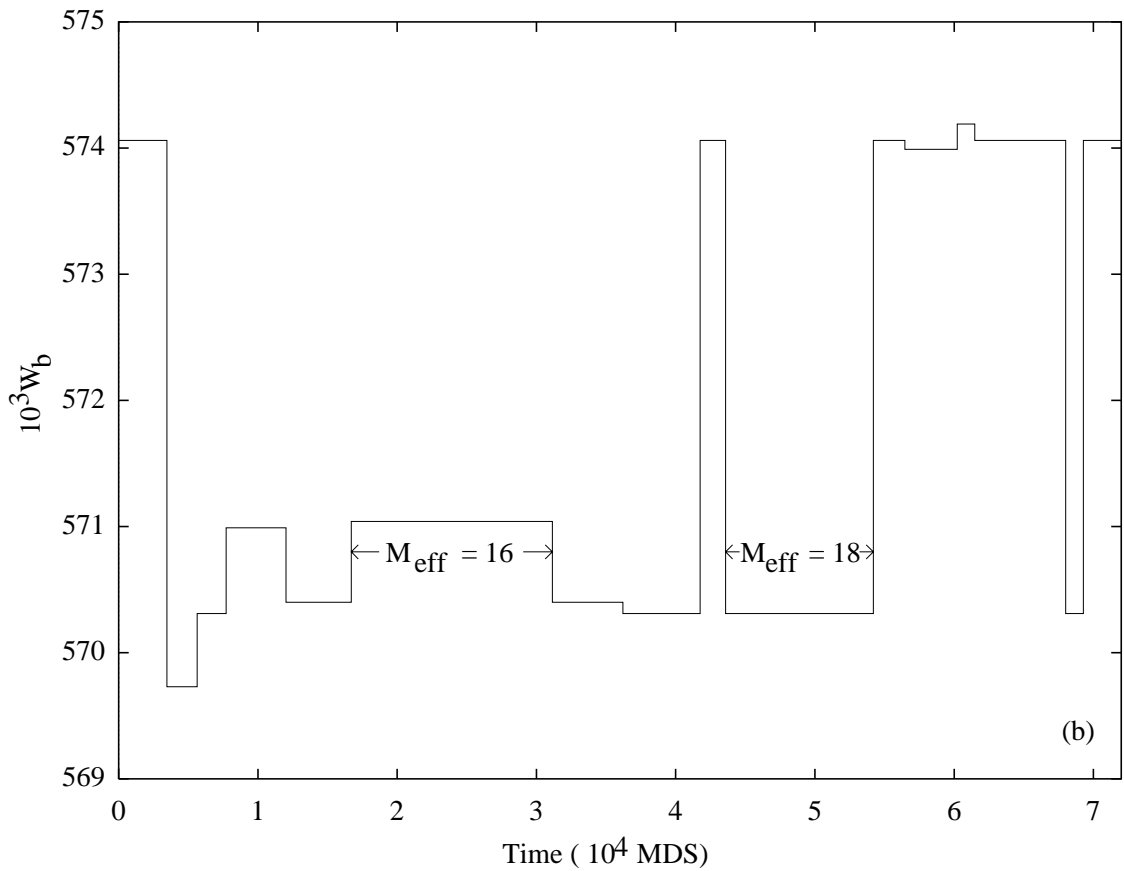
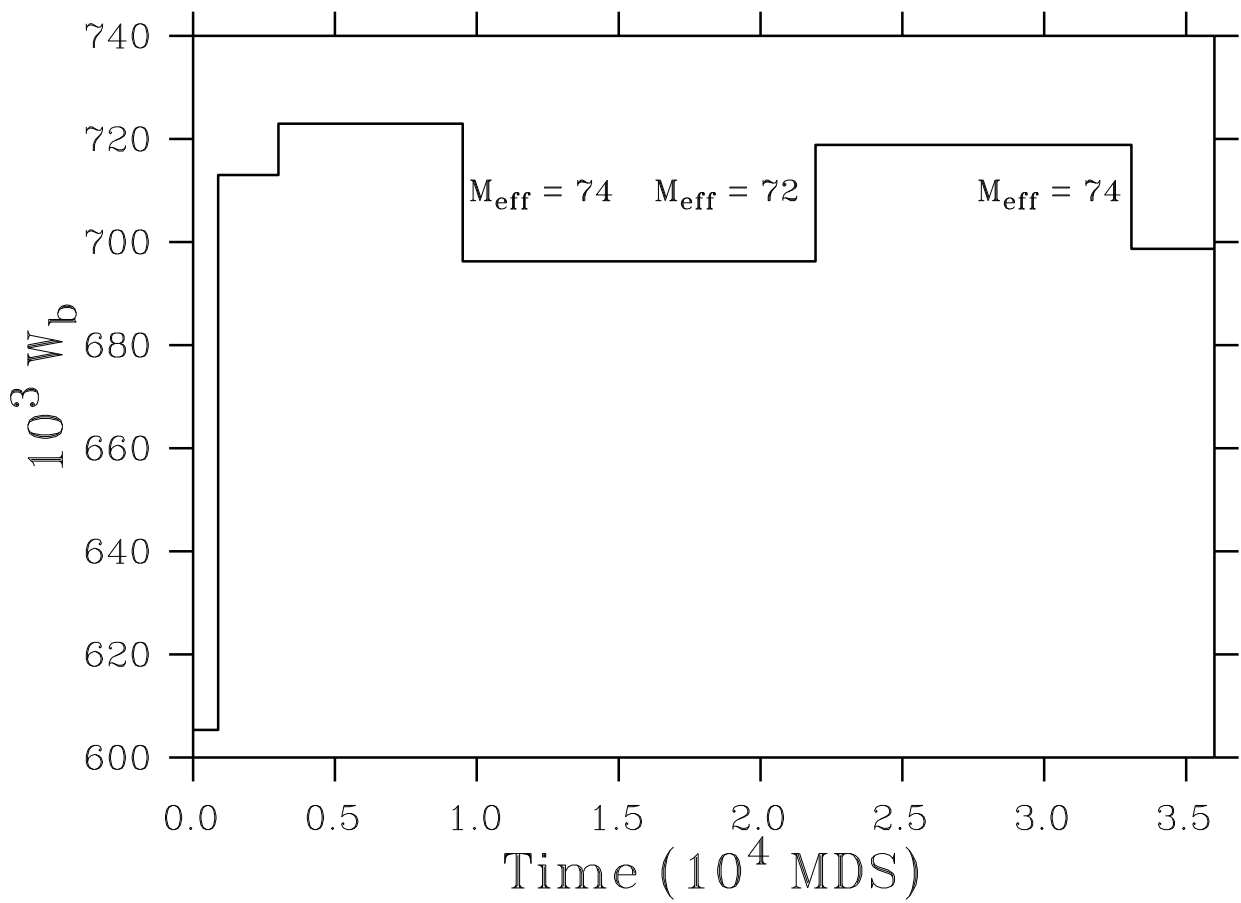
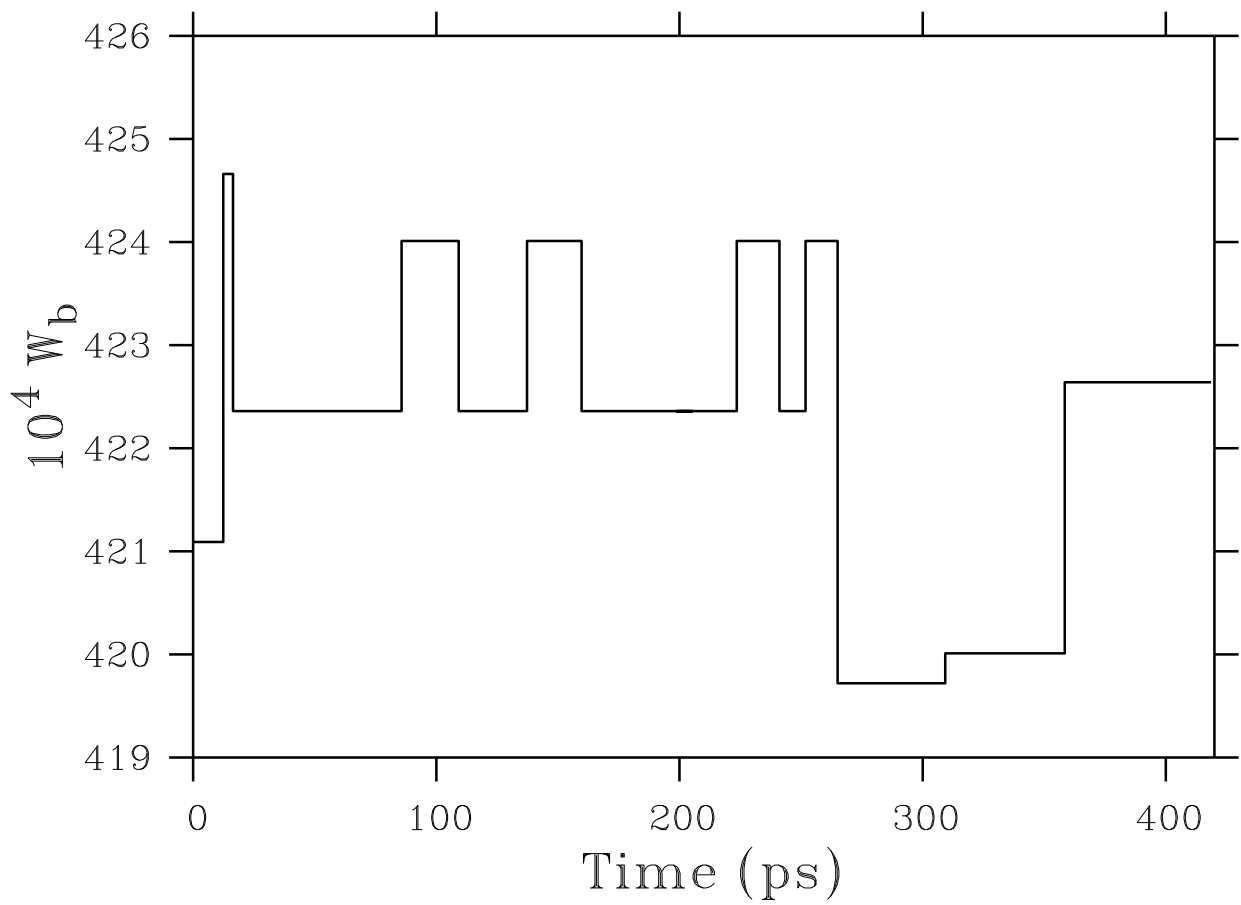
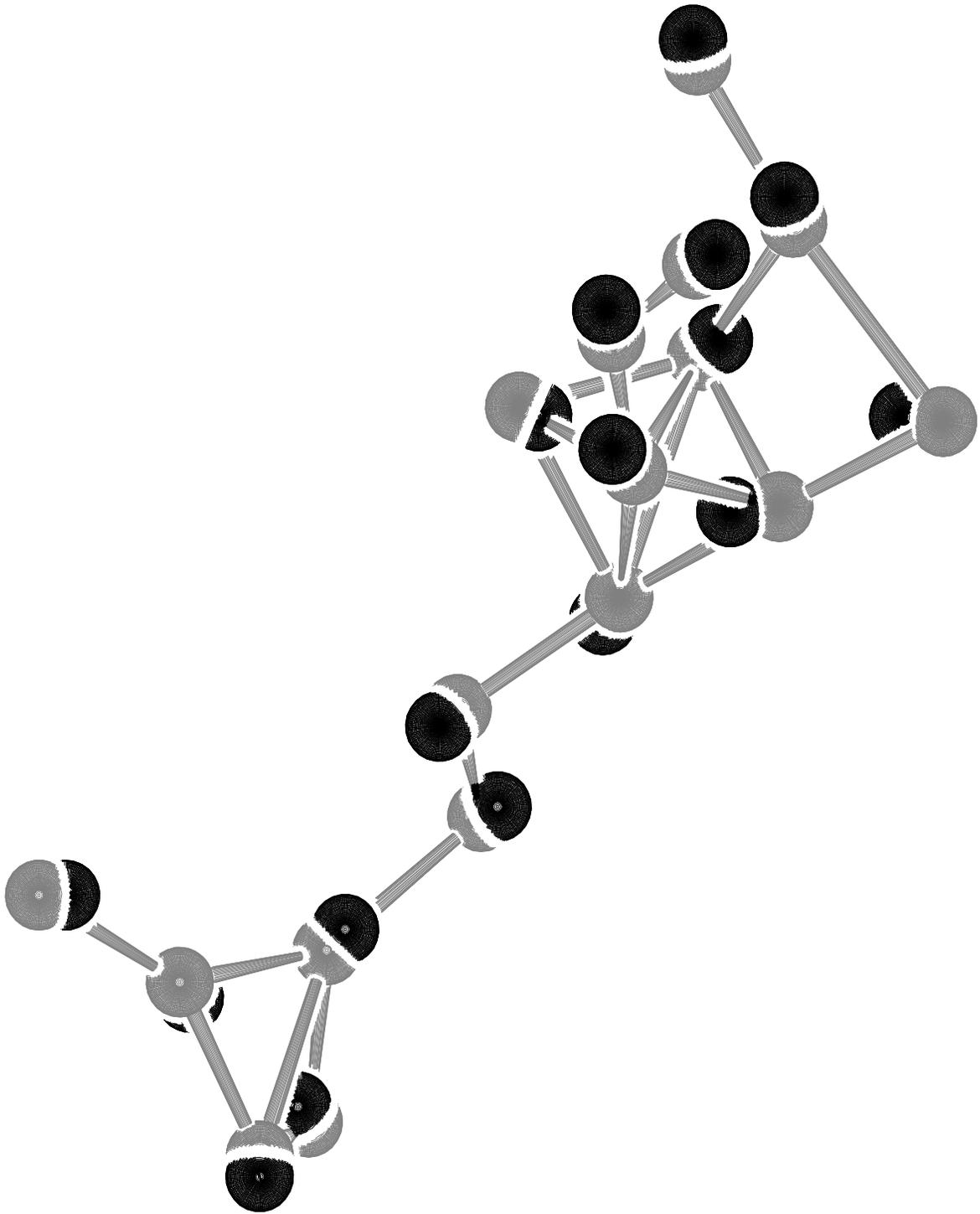
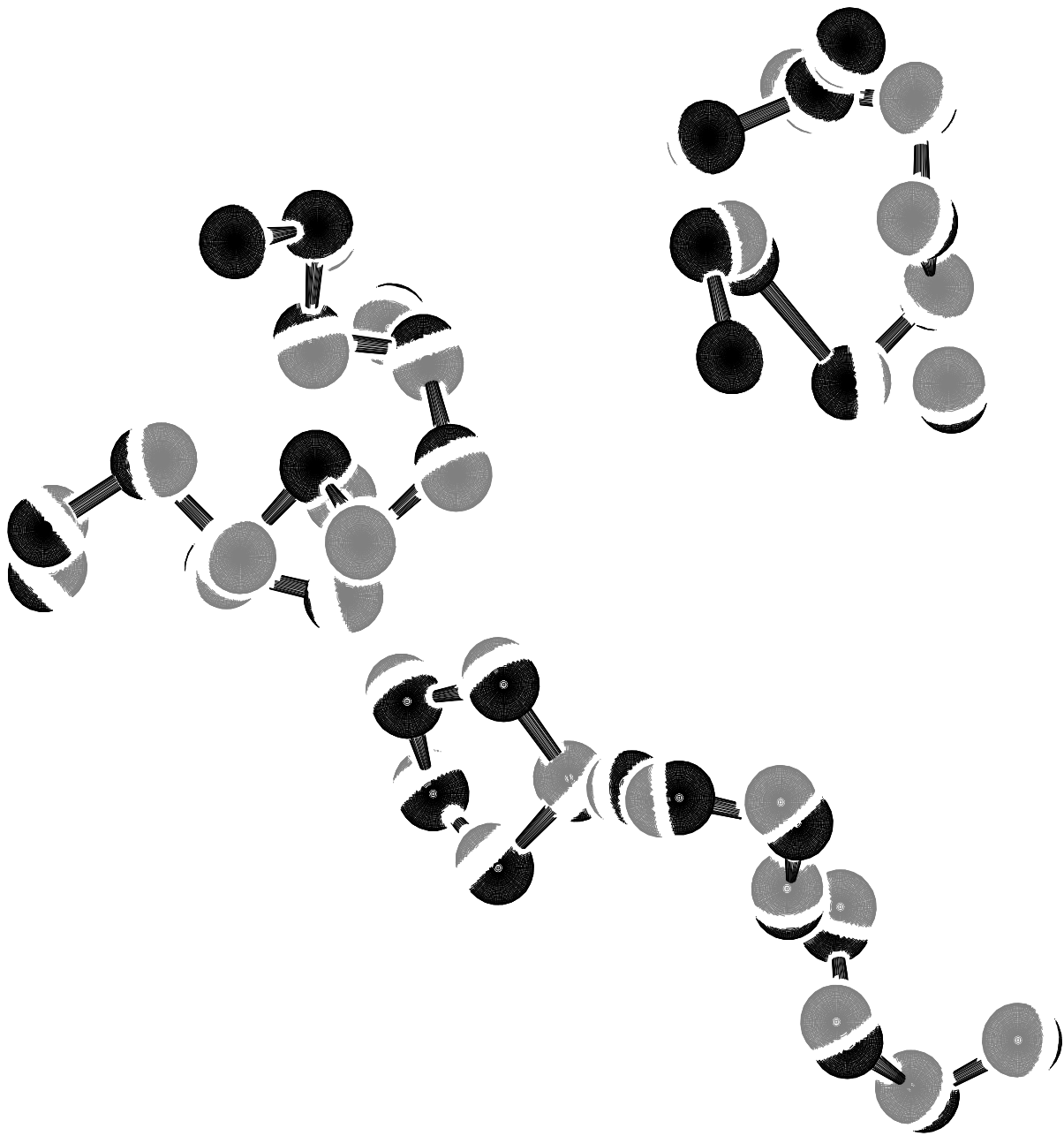
\end{document}